\documentclass{vgtc}                          

\ifpdf
  \pdfoutput=1\relax                   
  \usepackage{graphicx}                
  \DeclareGraphicsExtensions{.pdf,.png,.jpg,.jpeg} 
\else
  \usepackage{graphicx}                
  \DeclareGraphicsExtensions{.eps}     
\fi%

\graphicspath{{figures/}{pictures/}{images/}{./}} 

\usepackage{microtype}                 
\PassOptionsToPackage{warn}{textcomp}  
\usepackage{textcomp}                  
\usepackage{mathptmx}                  
\usepackage{times}                     
\usepackage{cite}                      
\usepackage{tabu}                      
\usepackage{booktabs}                  
\usepackage{amsmath}
\usepackage{amsfonts}
\newcommand{\ta}[1]{{\textcolor{red}{#1}}}
\usepackage{subcaption}
\usepackage{graphicx}
\usepackage{verbatim}
\usepackage{dblfloatfix}
\newcommand{\Rspace}{{\mathbb R}}
\usepackage{enumitem}
\setitemize{noitemsep,topsep=0pt,parsep=0pt,partopsep=0pt}

\usepackage{xcolor}

\newcommand\blfootnote[1]{%
  \begingroup
  \small
  \renewcommand\thefootnote{}\footnote{#1}%
  \addtocounter{footnote}{-1}%
  \endgroup
}

\onlineid{0}

\vgtccategory{Research}

\vgtcinsertpkg

\title{An Entropy-Based Test and Development Framework for Uncertainty Modeling in Level-Set Visualizations}

\author{Robert Sisneros\thanks{e-mail: sisneros@illinois.edu}\\ %
        \scriptsize National Center for Supercomputing Applications %
\and Tushar M. Athawale\thanks{e-mail:athawaletm@ornl.gov}\\ %
     \scriptsize Oak Ridge National Laboratory %
     \and David Pugmire\thanks{e-mail:pugmire@ornl.gov}\\ %
     \scriptsize Oak Ridge National Laboratory  %
     \and Kenneth Moreland\thanks{e-mail:morelandkd@ornl.gov}\\ %
     \scriptsize Oak Ridge National Laboratory }

\abstract{
We present a simple comparative framework for testing and developing uncertainty modeling in uncertain marching cubes implementations.
The selection of a model to represent the probability distribution of uncertain values directly influences the memory use, run time, and accuracy of an uncertainty visualization algorithm.
We use an entropy calculation directly on ensemble data to establish an expected result and then compare the entropy from various probability models, including uniform, Gaussian, histogram, and quantile models.
Our results verify that models matching the distribution of the ensemble indeed match the entropy.
We further show that fewer bins in nonparametric histogram models are more effective whereas large numbers of bins in quantile models approach data accuracy.
} 

 \CCScatlist{ 
   \CCScat{300}{Human-centered computing}{Visualization application domains}{Scientific Visualization}
 }

\begin{document}
\maketitle

\section{Introduction}
\label{sec:introduction}

Uncertainty is present in nearly all scientific data and arises from many factors including: sensor noise and tolerances, approximations in simulation models, resolutions of simulation grids, discretization, data reduction, etc.
The visualization of uncertainty is an important area of research to aid scientists to understand and trust their data~\cite{TA:Johnson:2003:nextStepVisErrors, TA:2004:Johnson:topScivis}.
To this end, research has been done to address uncertainty in different parts of the visualization pipeline. Many new algorithms have been devised to assess uncertainty in scalar field data. These algorithms include uncertainty analysis of level~sets~\cite{ TA:Whitaker:2013:contourBoxPlots, TA:Kai:2013:nonparametricIsoVis, TA:Athawale:2016:nonparametricIsosurfaces, TA:Athawale2019ProbAsympDecider}, direct volume rendering~\cite{TA:Lundstrom:2007:probabilisticAnimationMedicalVolRendering, TA:Fout:2012:fuzzyVolumeRendering,TA:Shusen:2012:GMMdvr, TA:Athawale:2021:nonparametricDVR}, and topology-based visualizations~\cite{TA:Gunther:2014:MandatoryCriticalPoints, TA:Favelier:2019:criticalPointVariabilityEnsembles, TA:Athawale:2022:uncertainMorseComplexes}. A few studies have investigated the uncertainty in visualizations of multivariate~\cite{bivariateFiberUncertainty,featureConfidenceLevelSets,scatterPlotUncertainty}, vector-field~\cite{TA:Lodha:1996:flowVisUncertainty, TA:Otto:2011:3dVectorFieldTopologyUncertainty, TA:Ferstl:2016:streamlineVariabilityVis, TA:Guo:2016:LyapunovUncertaintyUnsteadyFlow} and tensor-field~\cite{TA:Jones:2003:uncertaintyConeGlyphsTractography, TA:Jiao:2012:hardiDiffusionDataUncertaintyGlyph, TA:Siddiqui:2021:tensorImagingVisPipelineUncertainty} data. Overviews of challenges and state-of-the-art in uncertainty visualization are documented in multiple survey papers~\cite{TA:Brodlie:2012:RUDV,TA:Potter:2012:UQtaxonomy,TA:Bonneau:2014:StateOftheArtUQ, TA:Kamal:2021:UQvisSurvey}.



There are numerous models used to represent the distributions of uncertain data, and the choice of model is a critical factor for the time, memory, and effectiveness of an uncertain algorithm.
The effect of representations of uncertain data for visualization algorithms has received little attention.
Of particular importance is the memory cost for an uncertain representation, which becomes significant as the size of data continues to grow.
It is common, for example, for a simulation code running on a supercomputer to produce petabytes of data, or a large ensemble to contain hundreds of members. Likewise, advanced experimental facilities produce tera- and petabytes of data. At this scale, a compact representation of uncertainty is critical. Even the simplest representations of uncertainty, such as uniform or Gaussian distributions will add an additional factor of $2\times$ to the size of the data (min and max for uniform and mean and standard deviation for Gaussian).  Using a more accurate model such as a histogram will result in even larger space requirements.

In this short paper, we present a study where we compared both the accuracy (in terms of entropy) and the type and size of uncertainty representations on several ensemble data sets.
We consider the distribution using all ensemble members as the ground truth for the data.
We then compare this ground truth against uniform, Gaussian, histogram, and quantile models.
Our evaluation compares and contrasts the total entropy of each distribution model along with the representation costs.
We will present a simple entropy-based framework for the comparison and development of uncertainty models for visualization algorithms and describe multiple models.  We will conclude with a preliminary study comparing the accuracy of four different representations of uncertainty against our developed testing standard.

\section{Background and Related Works}
\label{background}


\begin{figure*}[tb]
     \begin{subfigure}[]{0.25\textwidth}
         \centering
         \includegraphics[width=\linewidth]{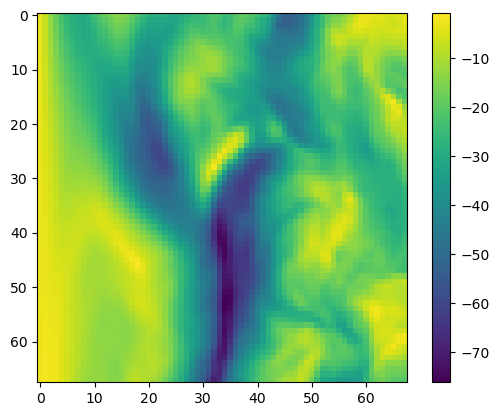}
         \caption{Original Dataset}
         \label{fig:original_wind}
     \end{subfigure}
     \hfill
     \begin{subfigure}[]{0.25\textwidth}
         \centering
         \includegraphics[width=\linewidth]{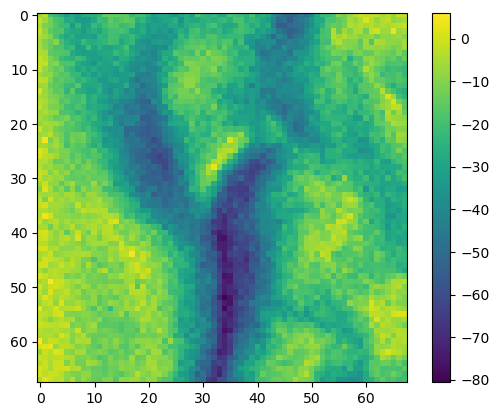}
         \caption{With Gaussian Noise}
         \label{fig:wind_gaussian}
     \end{subfigure}
     \hfill
     \begin{subfigure}[]{0.25\textwidth}
         \centering
         \includegraphics[width=\linewidth]{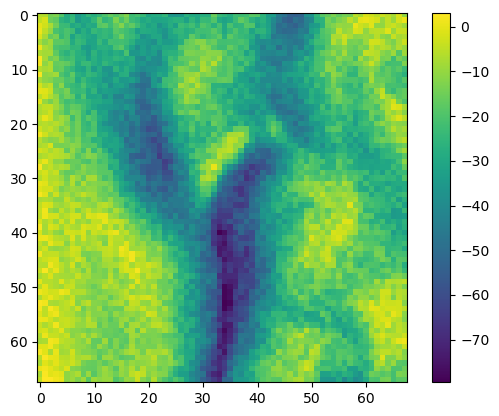}
         \caption{With Uniform Noise}
         \label{fig:wind_uniform}
     \end{subfigure}
        \caption{Renderings of the negative of the velocity magnitude of the first ensemble (of 15) of the wind dataset without and with noise.
        }
        \label{fig:wind_noise}
\end{figure*}

 For univarite function $f: \mathcal{D} \subset \Rspace^n \to \Rspace$ over domain $\mathcal{D}$, the level-set (L) for isovalue $k$ is given by $L = \{x|f(x)=k\}$. The marching squares (MS)/cubes (MC)~\cite{Lorensen:1987:MCA} is a fundamental visualization algorithm for level-set visualization of scientific data. This work leverages the entropy-based framework  proposed by Athawale et al.~\cite{marchingCubesUncertainty} to evaluate the accuracy and costs of various distribution models used in uncertainty visualization of level-sets. The previous work investigated the entropy of distance fields of isocontours with the aim to identify representative isocontours that maximize entropy~\cite{infoIsocontours}. The goal of this work is significantly different from the one for the work by Hazarika et al. in that we study the effect on entropy of marching cubes cases for different distribution models. The main idea of the entropy-based approach~\cite{marchingCubesUncertainty} is that the uncertainty propagation in level-set positions can be understood by deriving the probability distribution of $16$ topology cases in MS and $256$ topology cases in MC. Having derived the probability distribution of MS/MC topology cases, the entropy of the distribution can be computed and visualized to gain insight into level-set positions that are more or less sensitive to uncertainty in data. Generally speaking, more entropy means less confidence or higher uncertainty of level-set topology. We briefly describe this process of computing entropy/positional uncertainty of level-sets, as it is the basis for our evaluation of uncertainty models. 
\blfootnote{This manuscript has been authored by UT-Battelle, LLC under Contract No. DE-AC05-00OR22725 with the U.S. Department of Energy. The publisher, by accepting the article for publication, acknowledges that the U.S. Government retains a non-exclusive, paid up, irrevocable, world-wide license to publish or reproduce the published form of the manuscript, or allow others to do so, for U.S. Government purposes. The DOE will provide public access to these results in accordance with the DOE Public Access Plan (\url{http://energy.gov/downloads/doe-public-access-plan}).}

When data have uncertainty, we denote the data at cell vertices in a 2D case by random variable $D_{xy}$ (and $D_{xyz}$ in a 3D case).  Let $\text{pdf}_{D_{xy}}$ be the probability distribution at each vertex estimated from sample data. For the isovalue $k$, let $D^+_{xy} = Pr(D_{xy} \ge k)$, i.e., the probability of a cell vertex (x,y) attaining a positive vertex sign. Similarly, let $D^-_{xy} = Pr(D_{xy} < k)$. Having computed $D^+_{xy}$ and  $D^-_{xy}$ for each cell vertex, the probability for each of the $16$ MS topology cases can be computed per cell. For example, for the independent noise assumption, the probability of $D_{00}$, $D_{10}$, and $D_{11}$ being positive and $D_{01}$ being negative is equal to the product $D^+_{00} \cdot D^-_{01}\cdot D^+_{10} \cdot D^+_{11}$. 

Let $C$ denote a discrete random variable representing the $16$ MS topology cases, and $\text{pdf}_{C}(c)$ denote the topology case probability distribution.
The level-set entropy in a grid cell ($E$) then can be computed using the Shannon entropy~\cite{Shannon1948}, i.e., $E =-\sum_{c=1}^{c=16} Pr_{C}(c) log_2 Pr_{C}(c)$.
Visualization of $E$, therefore, provides an insight into confidence regarding level-set positions extracted from uncertain data.
In this paper, we analyze sensitivity of $E$ to uniform-, Gaussian-, histogram-, and quantile-based distribution models, and present important findings about comparisons of these distribution models.
This work is limited to analysis of only independent distribution models.

\begin{figure}[htb]
\centering
     \begin{subfigure}[]{0.3\textwidth}
         \centering
         \includegraphics[width=\linewidth]{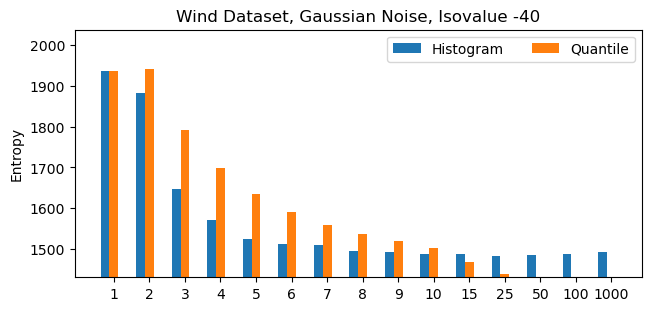}
         \caption{Gaussian noise}
         \label{fig:wind_entropy_gaussian}
     \end{subfigure}
     \hfill
     \begin{subfigure}[]{0.3\textwidth}
         \centering
         \includegraphics[width=\linewidth]{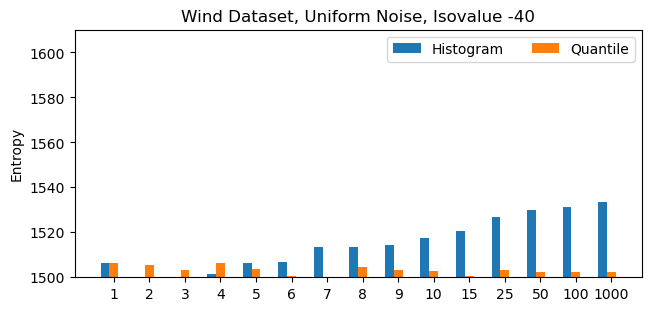}
         \caption{Uniform noise}
         \label{fig:wind_entropy_uniform}
     \end{subfigure}
        \caption{
          Total summed entropy from contouring the wind dataset using the histogram and quantile models containing between 1 and 1000 bins.
          The baseline of each chart is set at the target entropy of the full distribution (see \autoref{tab:windNoise}). Quantile models generally converge to the baseline entropy with an increase in the number of quantiles.
        }
        \label{fig:wind_entropy_noise}
\end{figure}

\begin{figure*}[htb]
     \begin{subfigure}[]{0.25\textwidth}
         \centering
         \includegraphics[width=\linewidth]{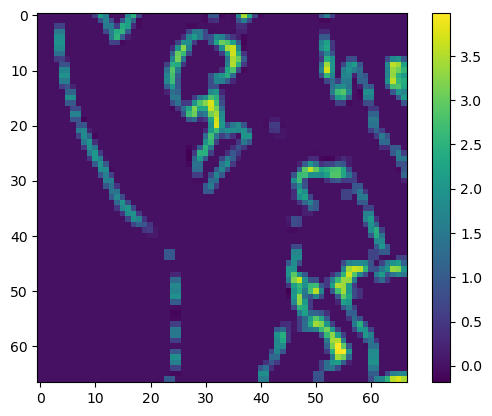}
         \caption{Isovalue -20}
         \label{fig:wind_iso20}
     \end{subfigure}
     \hfill
     \begin{subfigure}[]{0.25\textwidth}
         \centering
         \includegraphics[width=\linewidth]{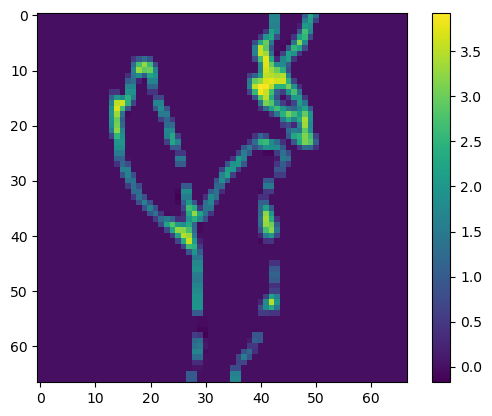}
         \caption{Isovalue -40}
         \label{fig:wind_iso40}
     \end{subfigure}
     \hfill
     \begin{subfigure}[]{0.25\textwidth}
         \centering
         \includegraphics[width=\linewidth]{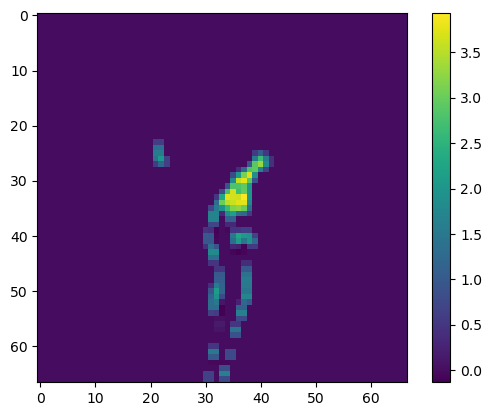}
         \caption{Isovalue -60}
         \label{fig:wind_iso60}
     \end{subfigure}
     \hfill
        \caption{
          Entropy of uncertain contours from the full distribution of the wind dataset.
        }
        \label{fig:wind_iso}
\end{figure*}

\begin{figure*}[t]
\centering
     \begin{subfigure}[]{0.3\textwidth}
         \centering
         \includegraphics[width=\linewidth]{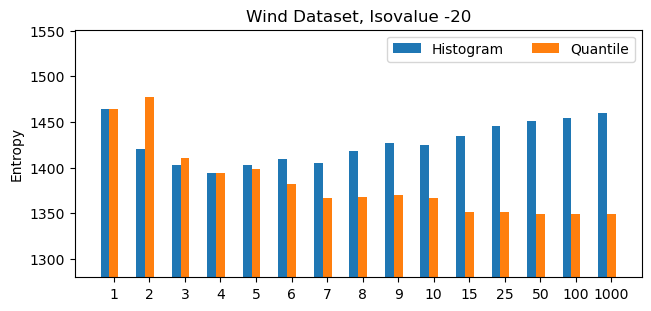}
         \caption{Isovalue -20}
         \label{fig:wind_iso20_bar}
     \end{subfigure}
     \hfill
     \begin{subfigure}[]{0.3\textwidth}
         \centering
         \includegraphics[width=\linewidth]{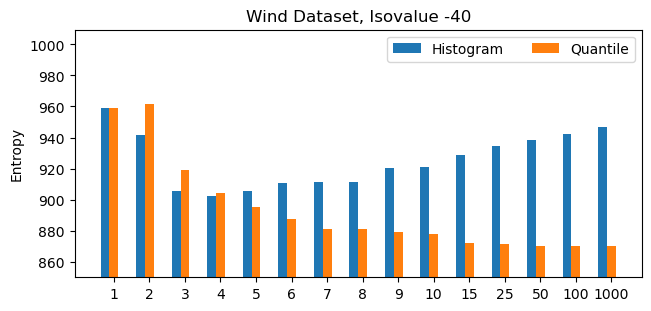}
         \caption{Isovalue -40}
         \label{fig:wind_iso40_bar}
     \end{subfigure}
     \hfill
     \begin{subfigure}[]{0.3\textwidth}
         \centering
         \includegraphics[width=\linewidth]{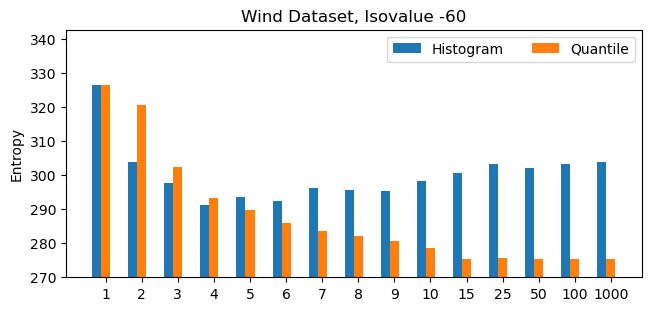}
         \caption{Isovalue -60}
         \label{fig:wind_iso60_bar}
     \end{subfigure}
        \caption{
          Total summed entropy from contouring the wind dataset using the histogram and quantile models containing between 1 and 1000 bins.
          The baseline of each chart is set at the target entropy of the full distribution (see \autoref{tab:wind}).
        }
        \label{fig:wind_charts}
\end{figure*}

\section{Evaluation Framework}
\label{framework}

\begin{table}[htb]
    \centering
    \small
    \begin{tabular}{c cc}
    \toprule
    & \multicolumn{2}{c}{Noise Distribution}\\
        Distribution & Gaussian & Uniform\\ 
        \midrule
        Full Dist. & 1438.71 & 1500.83 \\
        Uniform & 1936.24 & 1506.48 \\
        Gaussian & 1470.10 & 1435.39 \\
        Histogram & 1523.68 & 1506.47 \\
        Quantile & 1635.14 & 1503.46 \\
        \bottomrule
    \end{tabular}
    \caption{
      Total summed entropy from contouring the wind dataset at isovalue -40 with noise added.
      The values for the histogram and quantile distribution models are for the 5 bin case.
    }
    \label{tab:windNoise}
\end{table}

\begin{table}[b]
    \centering
    \small
    \begin{tabular}{c ccc }
    \toprule
    & & Isovalue & \\
        Distribution & -20 & -40 & -60\\ 
        \midrule
        Full Dist. & 1289.19 & 858.62 & 277.85 \\
        Uniform & 1463.65 & 959.14 & 326.29  \\
        Gaussian & 1389.77 & 892.08 &  281.96 \\
        Histogram & 1402.48 & 905.39 & 293.67  \\
        Quantile & 1398.74 & 895.24 &  289.81 \\
        \bottomrule
    \end{tabular}
    \caption{
      Total summed entropy from contouring the wind dataset under different distributions models.
      The values for the histogram and quantile distribution models are for the 5 bin case.
      }
    \label{tab:wind}
\end{table}

Our framework is for an uncertain MS/MC algorithm developed to generate level-sets from ensemble datasets.
Given such a dataset, we treat the ensembles as samples from some distribution and first employ a routine to provide a reduced representative model of this distribution.
This model is then used to calculate the probabilities of each topological MS/MC case.
As described previously, large data sets benefit greatly from a representation of a reduced size.

In previous work~\cite{TA:Athawale:2016:nonparametricIsosurfaces}, level-set probabilities were estimated using a simple histogram model that provided better precision in the contour estimation.
This precision comes at the expense of additional memory costs over simpler models such as simply assuming that a uniform distribution~\cite{AthawaleEntezari2013} exists between the minimum and maximum ensemble values.
The histogram model requires a value for each bin and two additional values for the bin range.
In contrast, only two values are required for the uniform model, a min and a max, and for the Gaussian model, a mean and standard deviation.
For a given ensemble and with no a priori knowledge, how do we choose the best model or even compare two models?
When deploying the histogram model, how do we select the number of bins to use? There is no work, to our knowledge, that studies trade offs between uncertainty models and quality of uncertainty visualization. 

As discussed in Sec.~\ref{background}, entropy is a direct measure of how uncertain level-set positions are in the domain. 
Although lower entropy corresponds to ``certainty'' in the algorithm, this does not mean lower values are more correct.
There are many uninteresting or even incorrect results that could be quite certain.
For instance, simply using a single ensemble member as a representative, taking the average across all ensembles results, or selection of an isovalue outside extents all result in zero entropy.
These are trivial or known for poor performance~\cite{visWithoutMean}.
That is, an improvement of an algorithm should reduce its entropy, but minimization of entropy is not an appropriate design standard.
As we will show below, we first estimate a target entropy then evaluate performance via comparison to that. 

To use entropy as a metric for comparing performance, we need a baseline of the best we can expect to perform.
With no memory constraints, we could simply use all the ensemble values by treating the probability density function as impulses of the ensemble values and directly calculating the probability an ensemble member is above or below an isovalue.
We use this as a standard for comparison and a benchmark for analysis. Note that our ensemble-driven approach can be treated as a baseline assuming that the sampled data is representative of underlying distribution.In the next Section, we will detail our early tests using this framework on two datasets.
\begin{table}[htb]
    \centering
    \small 
    \begin{tabular}{c ccc}
    \toprule
    & \multicolumn{2}{c}{2D Slice} & 3D Subset\\
        Distribution & Isovalue 0.15 &  Isovalue 0.5 & Isovalue 0.15\\ 
        \midrule
        Full Dist. & 32185.87 & 9354.90 & 218368.58 \\
        Uniform & 50373.50 & 15932.14 & 339110.23 \\
        Gaussian & 36532.71 & 11687.51 & 272221.33 \\
        Histogram & 39612.18 & 12495.03 & 279988.95 \\
        Quantile & 40536.70 & 12611.23 & 284453.83 \\
        \bottomrule
    \end{tabular}
    \caption{
      Total summed entropy from the Red Sea dataset.
      The values for the histogram and quantile distribution models are for the 5 bin case.
    }
    \label{tab:redsea}
\end{table}

\section{Results and Discussion}
\label{results}

\begin{figure*}[tb]
     \begin{subfigure}[]{0.25\textwidth}
         \centering
         \includegraphics[width=\linewidth]{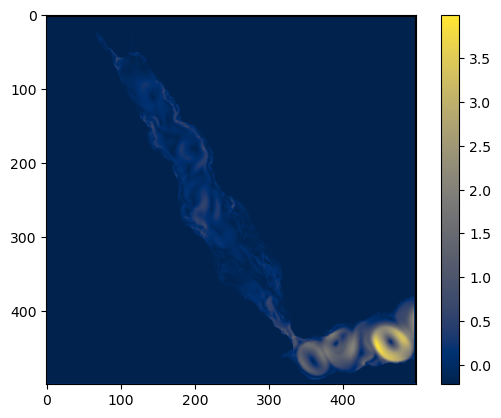}
         \caption{Original Dataset}
         \label{fig:original_redsea}
     \end{subfigure}
     \hfill
     \begin{subfigure}[]{0.25\textwidth}
         \centering
         \includegraphics[width=\linewidth]{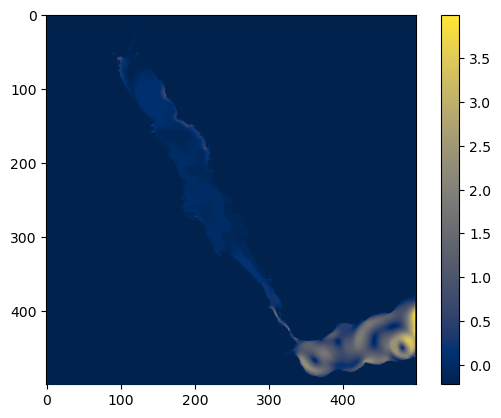}
         \caption{Slice Dataset}
         \label{fig:redsea_slice}
     \end{subfigure}
     \hfill
     \begin{subfigure}[]{0.25\textwidth}
         \centering
         \includegraphics[width=\linewidth]{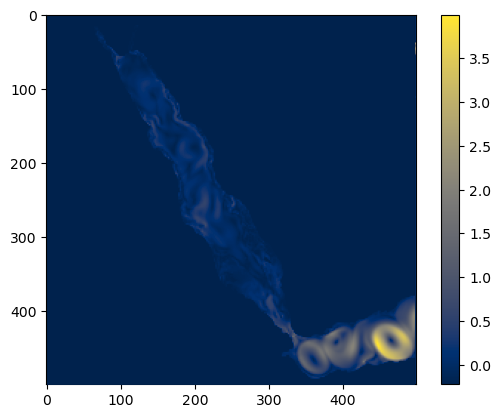}
         \caption{Subsampled Dataset}
         \label{fig:redsea_subsample}
     \end{subfigure}
        \caption{Velocity magnitude of the Red Sea dataset.}
        \label{fig:redsea}
\end{figure*}

\begin{figure*}
\centering
     \begin{subfigure}[]{0.3\textwidth}
         \centering
         \includegraphics[width=\linewidth]{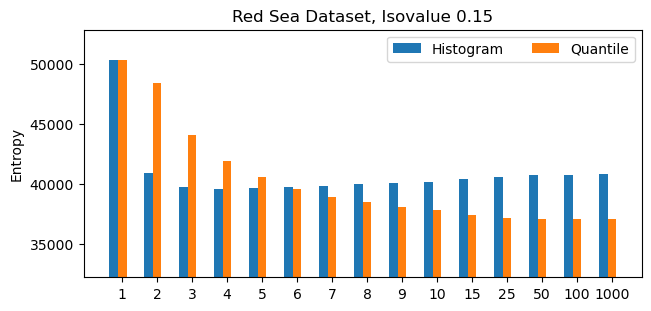}
         \caption{2D slice, Isovalue 0.15}
         \label{fig:rsliceBins_iso_point15}
     \end{subfigure}
     \hfill
     \begin{subfigure}[]{0.3\textwidth}
         \centering
         \includegraphics[width=\linewidth]{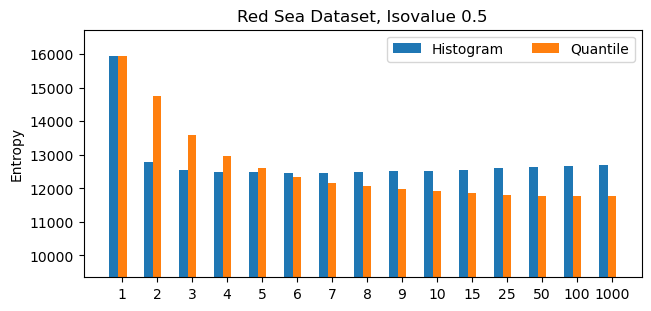}
         \caption{2D slice, Isovalue 0.5}
         \label{fig:rsliceBins_iso_point5}
     \end{subfigure}
     \hfill
     \begin{subfigure}[]{0.3\textwidth}
         \centering
         \includegraphics[width=\linewidth]{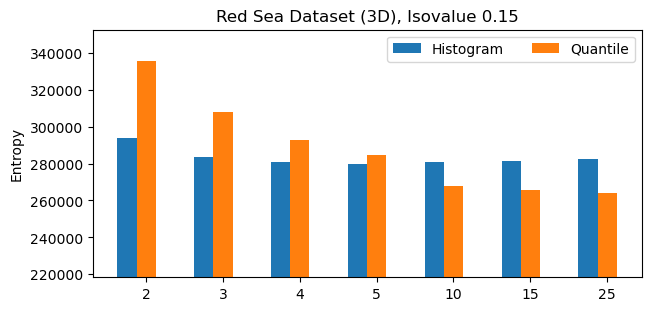}
         \caption{3D subsampling, Isovalue 0.15}
     \end{subfigure}

    \caption{
        Total summed entropy from contouring the Red Sea dataset using the histogram and quantile models containing between 1 and 1000 bins.
        The baseline of each chart is set at the target entropy of the full distribution (see \autoref{tab:redsea}).
    }
        \label{fig:redsea_charts}
\end{figure*}

We used two data sets for this study.
The \textbf{Red Sea} Dataset~\cite{Zhan2014} is publicly accessible through the IEEE SciVis contest 2020. It contains the velocity of water at a resolution of $500\times500\times50$ on a uniform grid with $20$ ensemble members.
The \textbf{Wind} Dataset~\cite{vitart2017subseasonal} from the IRI/LDEO Climate Data Library contains the wind velocity at a resolution of  $68\times68$ on a uniform grid with $15$ ensemble members.

\subsection{Wind Dataset Results}

Our preliminary testing was done with the wind dataset.  \autoref{tab:wind} shows our early test results for three isovalues on the Wind ensemble dataset, shown in \autoref{fig:original_wind}.  Plots showing entropic areas of interest corresponding to each isovalue are in \autoref{fig:wind_iso}.   

Not shown in \autoref{tab:wind} is the effect the number of bins has on the histogram and quantile models.
To study this, entropy is reported across all tested bin counts in the charts of \autoref{fig:wind_charts}.
We see the quantile model is monotonically decreasing, but requires more bins than the histogram to effectively reduce entropy.
Note that with a single bin the histogram and quantile are the same as each other and the same as the uniform model.
These are included for reference. 

With the framework in place, we want to evaluate whether our target entropy makes sense.
To do this, we developed the following test:  take a representative ensemble, create an ensemble of 50 members via addition of random Gaussian noise, and compare the entropy of the random ensemble with the uncertainty models that do and do not match the noise.
The same is repeated for uniform noise.
Representative examples of pre- and post-noise data are in \autoref{fig:wind_noise}.
Results of these tests are in \autoref{tab:windNoise} with bins reported in \autoref{fig:wind_entropy_noise}.
The full distribution model is indeed aligned with an estimate of the correct model. 

One oddity of these results is that for the uniform noise distribution the Gaussian distribution model had a lower entropy than the full distribution.
This case highlights the subtlety of using entropy as a metric.
The Gaussian acts to smooth the ensembles toward the mean, which practically speaking, can lead to a more pleasing visual result, but this is at the expense of removing entropy that actually should be considered.
The simple uniform tests also highlight the negative impact for over-binning in the histogram model. 

\subsection{Red Sea Results}
\begin{figure}[htb]
    \begin{subfigure}[]{0.24\textwidth}
         \centering
         \includegraphics[width=\linewidth]{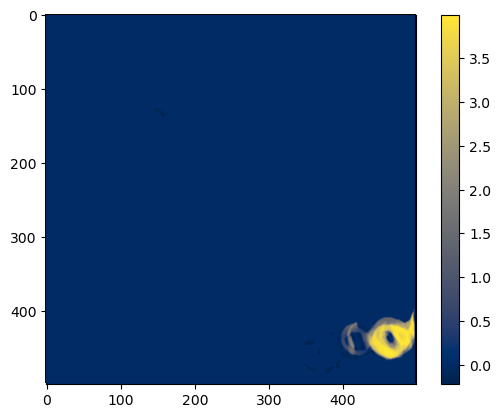}
         \caption{2D Slice, Isovalue 0.5}
         \label{fig:rslice_iso_point5}
     \end{subfigure}
     \hfill
     \begin{subfigure}[]{0.24\textwidth}
         \centering
         \includegraphics[width=\linewidth]{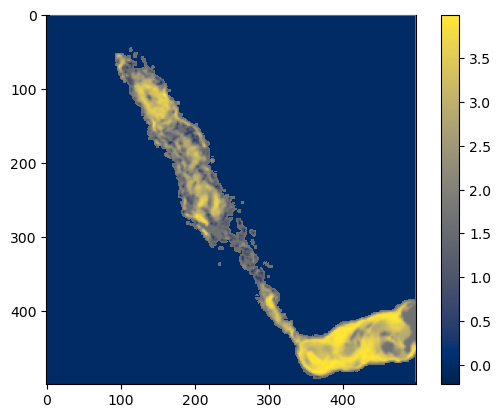}
         \caption{Subsampled Data, Isovalue 0.15}
         \label{fig:redsea_iso_subsampled}
     \end{subfigure}
     \hfill

        \caption{
            Entropy of uncertain contours for the Red Sea dataset.
            Isovalues 0.15 and 0.5 (a) were used for the 2D tests. Only isovalue 0.15 was used for the 3D tests, shown for subsampled data (b).
        }
        \label{fig:redsea_iso}
\end{figure}

 In this Section, we use the Red Sea data set to validate our testing framework on larger, more complicated ensemble data.
 Each test on the full dataset (\autoref{fig:original_redsea}) took over four hours, which made it difficult to explore a large space of options.
 To accelerate the analysis, we first operated on 2D slices of the dataset  (\autoref{fig:redsea_slice}) and subsampled (each dimension by 2) 3D versions of the dataset (\autoref{fig:redsea_subsample}).
 Insight derived from the analysis of these smaller datasets helped steer the more expensive analysis of the full 3D datasets.
\autoref{fig:redsea_iso} shows representative isovalues used in testing.  All scenarios used isovalue 0.15, the 2D tests also used 0.5.
 
The Red Sea 2D slice tests are reported in \autoref{tab:redsea} and \autoref{fig:redsea_charts}.
The increased entropy from histogram over-binning is less pronounced here, which meets our expectations better.
Surprisingly, the areas where histogram and quantile models ``cross over'' is similar, but the much lower entropy with a small bin count is interesting. 



During our tests, we also collected execution times and found that overall execution times were similar for all models, with the quantile model being consistently slightly more computationally expensive than the histogram (less than 5\%).  We did see an uptick in overhead for large datasets when bin sizes were larger (100 or greater).  

The largest scale 3D test results are well aligned with our smaller, sampled tests, suggesting potential for doing so as a standard. Overall, the quantile model is interesting particularly as a benchmark to understand the trade offs of other models.
In more limited situations in practice, the quality of modeling may not be worth additional costs.
In such situations a histogram model with few bins provides more accurate models.

\section{Conclusion and Future Directions}

We show a simple, analytical, and comparative framework for uncertain MS/MC implementations and the representation of uncertainty used therein.
In this framework, the uncertain contours of an initial ensemble are evaluated, and the entropy of the contours is used as a target for the uncertain metrics.
From there, uncertainty models are derived from the ensemble, uncertain contours are computed, and the contour entropy of ensemble and model are compared.  First, the results in this paper were generated on reduced datasets.  While early tests progressed as expected while scaling, we intend to further analyze reduced-scale testing and/or scale tests in future work. 

In our experiments, we found that a uniform distribution model was further from the target entropy than we expected.
Also, while the Gaussian model appears to outperform the histogram model by generally achieving entropy closer to the target, we have shown there are distributions for which the model, perhaps incorrectly, reports lower entropy.
In practice, this could be beneficial as uncertainty must be addressed somewhere in the pipeline and Gaussian smoothing is a routine deployment.

Finally, in light of the potential for reducing storage and memory costs, we plan to explore the development of adaptive methods for representing uncertainty. This adaptive method could use different models within a data set for a more space-efficient representation of the uncertainty.  In this way, in regions where uncertainty is nearly the same, a uniform model can be used to accurately represent (within some tolerance) the underlying data. In regions where the uncertainty varies, an appropriate model can be used for representation. In this way, we can spend storage in a much more cost-efficient manner.

\acknowledgments{
This work was supported in part by the U.S. Department of Energy (DOE) RAPIDS-2 SciDAC project under contract number DE-AC0500OR22725.}
\bibliographystyle{abbrv-doi}

\bibliography{template}

\begin{thebibliography}{10}

\bibitem{AthawaleEntezari2013}
T.~M. Athawale and A.~Entezari.
\newblock Uncertainty quantification in linear interpolation for isosurface extraction.
\newblock {\em IEEE Transactions on Visualization and Computer Graphics}, 19(12):2723--2732, 2013.

\bibitem{TA:Athawale2019ProbAsympDecider}
T.~M. Athawale and C.~R. Johnson.
\newblock Probabilistic asymptotic decider for topological ambiguity resolution in level-set extraction for uncertain {2D} data.
\newblock {\em IEEE Transactions on Visualization and Computer Graphics}, 25(1):1163--1172, Jan. 2019. doi: {{%
10\hspace{.1pt}\discretionary{.}{%
}{.}\hspace{.4pt}1109\discretionary{/}{%
}{/}TVCG\hspace{.1pt}\discretionary{.}{%
}{.}\hspace{.4pt}2018\hspace{.1pt}\discretionary{.}{%
}{.}\hspace{.4pt}2864505}}


\bibitem{bivariateFiberUncertainty}
T.~M. Athawale, C.~R. Johnson, S.~Sane, and D.~Pugmire.
\newblock Fiber uncertainty visualization for bivariate data with parametric and nonparametric noise models.
\newblock {\em IEEE Transactions on Visualization and Computer Graphics}, 29(1):613--623, 2023. doi: {{%
10\hspace{.1pt}\discretionary{.}{%
}{.}\hspace{.4pt}1109\discretionary{/}{%
}{/}TVCG\hspace{.1pt}\discretionary{.}{%
}{.}\hspace{.4pt}2022\hspace{.1pt}\discretionary{.}{%
}{.}\hspace{.4pt}3209424}}


\bibitem{TA:Athawale:2021:nonparametricDVR}
T.~M. Athawale, B.~Ma, E.~Sakhaee, C.~R. Johnson, and A.~Entezari.
\newblock Direct volume rendering with nonparametric models of uncertainty.
\newblock {\em IEEE Transactions on Visualization and Computer Graphics}, 27(2):1797--1807, Feb. 2021. doi: {{%
10\hspace{.1pt}\discretionary{.}{%
}{.}\hspace{.4pt}1109\discretionary{/}{%
}{/}TVCG\hspace{.1pt}\discretionary{.}{%
}{.}\hspace{.4pt}2020\hspace{.1pt}\discretionary{.}{%
}{.}\hspace{.4pt}3030394}}


\bibitem{TA:Athawale:2022:uncertainMorseComplexes}
T.~M. Athawale, D.~Maljovec, L.~Yan, C.~R. Johnson, V.~Pascucci, and B.~Wang.
\newblock Uncertainty visualization of 2{D} {M}orse complex ensembles using statistical summary maps.
\newblock {\em IEEE Transactions on Visualization and Computer Graphics}, 28(4):1955--1966, Apr. 2022. doi: {{%
10\hspace{.1pt}\discretionary{.}{%
}{.}\hspace{.4pt}1109\discretionary{/}{%
}{/}TVCG\hspace{.1pt}\discretionary{.}{%
}{.}\hspace{.4pt}2020\hspace{.1pt}\discretionary{.}{%
}{.}\hspace{.4pt}3022359}}


\bibitem{TA:Athawale:2016:nonparametricIsosurfaces}
T.~M. Athawale, E.~Sakhaee, and A.~Entezari.
\newblock Isosurface visualization of data with nonparametric models for uncertainty.
\newblock {\em IEEE Transactions on Visualization and Computer Graphics}, 19(12):2723--2732, Jan. 2016. doi: {{%
10\hspace{.1pt}\discretionary{.}{%
}{.}\hspace{.4pt}1109\discretionary{/}{%
}{/}TVCG\hspace{.1pt}\discretionary{.}{%
}{.}\hspace{.4pt}2015\hspace{.1pt}\discretionary{.}{%
}{.}\hspace{.4pt}2467958}}


\bibitem{marchingCubesUncertainty}
T.~M. Athawale, S.~Sane, and C.~R. Johnson.
\newblock Uncertainty visualization of the marching squares and marching cubes topology cases.
\newblock In {\em 2021 IEEE Visualization Conference (VIS)}, pp. 106--110, 2021. doi: {{%
10\hspace{.1pt}\discretionary{.}{%
}{.}\hspace{.4pt}1109\discretionary{/}{%
}{/}VIS49827\hspace{.1pt}\discretionary{.}{%
}{.}\hspace{.4pt}2021\hspace{.1pt}\discretionary{.}{%
}{.}\hspace{.4pt}9623267}}


\bibitem{TA:Bonneau:2014:StateOftheArtUQ}
G.~Bonneau, H.~Hege, C.~R. Johnson, M.~Oliveira, K.~Potter, P.~Rheingans, and T.~Schultz.
\newblock Overview and state-of-the-art of uncertainty visualization.
\newblock In M.~Chen, H.~Hagen, C.~Hansen, C.~R. Johnson, and A.~Kauffman, eds., {\em Scientific Visualization: Uncertainty, Multifield, Biomedical, and Scalable Visualization}, pp. 3--27. Springer London, 2014. doi: {{%
10\hspace{.1pt}\discretionary{.}{%
}{.}\hspace{.4pt}1007\discretionary{/}{%
}{/}978\discretionary{%
}{-}{-}1\discretionary{%
}{-}{-}4471\discretionary{%
}{-}{-}6497\discretionary{%
}{-}{-}5\_1}}


\bibitem{TA:Brodlie:2012:RUDV}
K.~Brodlie, R.~A. Osorio, and A.~Lopes.
\newblock A review of uncertainty in data visualization.
\newblock In J.~Dill, R.~Earnshaw, D.~Kasik, J.~Vince, and P.~C. Wong, eds., {\em Expanding the Frontiers of Visual Analytics and Visualization}, pp. 81--109. Springer Verlag London, 2012.

\bibitem{TA:Favelier:2019:criticalPointVariabilityEnsembles}
G.~Favelier, N.~Faraj, B.~Summa, and J.~Tierny.
\newblock Persistence atlas for critical point variability in ensembles.
\newblock {\em IEEE Transactions on Visualization and Computer Graphics}, 25(1):1152--1162, Jan. 2019. doi: {{%
10\hspace{.1pt}\discretionary{.}{%
}{.}\hspace{.4pt}1109\discretionary{/}{%
}{/}TVCG\hspace{.1pt}\discretionary{.}{%
}{.}\hspace{.4pt}2018\hspace{.1pt}\discretionary{.}{%
}{.}\hspace{.4pt}2864432}}


\bibitem{TA:Ferstl:2016:streamlineVariabilityVis}
F.~Ferstl, K.~Bürger, and R.~Westermann.
\newblock Streamline variability plots for characterizing the uncertainty in vector field ensembles.
\newblock {\em IEEE Transactions on Visualization and Computer Graphics}, 22(1):767--776, Jan. 2016. doi: {{%
10\hspace{.1pt}\discretionary{.}{%
}{.}\hspace{.4pt}1109\discretionary{/}{%
}{/}TVCG\hspace{.1pt}\discretionary{.}{%
}{.}\hspace{.4pt}2015\hspace{.1pt}\discretionary{.}{%
}{.}\hspace{.4pt}2467204}}


\bibitem{TA:Fout:2012:fuzzyVolumeRendering}
N.~Fout and K.-L. Ma.
\newblock Fuzzy volume rendering.
\newblock {\em IEEE Transactions on Visualization and Computer Graphics}, 18(12):2335--2344, Dec. 2012. doi: {{%
10\hspace{.1pt}\discretionary{.}{%
}{.}\hspace{.4pt}1109\discretionary{/}{%
}{/}TVCG\hspace{.1pt}\discretionary{.}{%
}{.}\hspace{.4pt}2012\hspace{.1pt}\discretionary{.}{%
}{.}\hspace{.4pt}227}}


\bibitem{TA:Gunther:2014:MandatoryCriticalPoints}
D.~G\"{u}nther, J.~Salmon, and J.~Tierny.
\newblock Mandatory critical points of 2{D} uncertain scalar fields.
\newblock {\em Computer Graphics Forum}, 33(3):31--40, July 2014. doi: {{%
10\hspace{.1pt}\discretionary{.}{%
}{.}\hspace{.4pt}1111\discretionary{/}{%
}{/}cgf\hspace{.1pt}\discretionary{.}{%
}{.}\hspace{.4pt}12359}}


\bibitem{TA:Guo:2016:LyapunovUncertaintyUnsteadyFlow}
H.~Guo, W.~He, T.~Peterka, H.-W. Shen, S.~M. Collis, and J.~J. Helmus.
\newblock Finite-time {L}yapunov exponents and {L}agrangian coherent structures in uncertain unsteady flows.
\newblock {\em IEEE Transactions on Visualization and Computer Graphics}, 22(6):1672--1682, June 2016. doi: {{%
10\hspace{.1pt}\discretionary{.}{%
}{.}\hspace{.4pt}1109\discretionary{/}{%
}{/}TVCG\hspace{.1pt}\discretionary{.}{%
}{.}\hspace{.4pt}2016\hspace{.1pt}\discretionary{.}{%
}{.}\hspace{.4pt}2534560}}


\bibitem{infoIsocontours}
S.~Hazarika, A.~Biswas, S.~Dutta, and H.-W. Shen.
\newblock Information guided exploration of scalar values and isocontours in ensemble datasets.
\newblock {\em Entropy}, 20(7), 2018. doi: {{%
10\hspace{.1pt}\discretionary{.}{%
}{.}\hspace{.4pt}3390\discretionary{/}{%
}{/}e20070540}}


\bibitem{TA:Jiao:2012:hardiDiffusionDataUncertaintyGlyph}
F.~Jiao, J.~M. Phillips, Y.~Gur, and C.~R. Johnson.
\newblock Uncertainty visualization in {HARDI} based on ensembles of {ODF}s.
\newblock In {\em 2012 IEEE Pacific Visualization Symposium}, pp. 193--200, Feb.-Mar. 2012. doi: {{%
10\hspace{.1pt}\discretionary{.}{%
}{.}\hspace{.4pt}1109\discretionary{/}{%
}{/}PacificVis\hspace{.1pt}\discretionary{.}{%
}{.}\hspace{.4pt}2012\hspace{.1pt}\discretionary{.}{%
}{.}\hspace{.4pt}6183591}}


\bibitem{TA:2004:Johnson:topScivis}
C.~R. Johnson.
\newblock Top scientific visualization research problems.
\newblock {\em IEEE Computer Graphics and Applications}, 24(4):13--17, 2004. doi: {{%
10\hspace{.1pt}\discretionary{.}{%
}{.}\hspace{.4pt}1109\discretionary{/}{%
}{/}MCG\hspace{.1pt}\discretionary{.}{%
}{.}\hspace{.4pt}2004\hspace{.1pt}\discretionary{.}{%
}{.}\hspace{.4pt}20}}


\bibitem{TA:Johnson:2003:nextStepVisErrors}
C.~R. Johnson and A.~R. Sanderson.
\newblock A next step: Visualizing errors and uncertainty.
\newblock {\em IEEE Computer Graphics and Applications}, 23(5):6--10, Sept.-Oct. 2003. doi: {{%
10\hspace{.1pt}\discretionary{.}{%
}{.}\hspace{.4pt}1109\discretionary{/}{%
}{/}MCG\hspace{.1pt}\discretionary{.}{%
}{.}\hspace{.4pt}2003\hspace{.1pt}\discretionary{.}{%
}{.}\hspace{.4pt}1231171}}


\bibitem{TA:Jones:2003:uncertaintyConeGlyphsTractography}
D.~K. Jones.
\newblock Determining and visualizing uncertainty in estimates of fiber orientation from diffusion tensor {MRI}.
\newblock {\em Magnetic Resonance in Medicine}, 49(1):7--12, Dec. 2002. doi: {{%
10\hspace{.1pt}\discretionary{.}{%
}{.}\hspace{.4pt}1002\discretionary{/}{%
}{/}mrm\hspace{.1pt}\discretionary{.}{%
}{.}\hspace{.4pt}10331}}


\bibitem{TA:Kamal:2021:UQvisSurvey}
A.~Kamal, P.~Dhakal, A.~Y. Javaid, V.~K. Devabhaktuni, D.~Kaur, J.~Zaientz, and R.~Marinier.
\newblock Recent advances and challenges in uncertainty visualization: a survey.
\newblock {\em Journal of Visualization}, 24(5):861--890, May 2021. doi: {{%
10\hspace{.1pt}\discretionary{.}{%
}{.}\hspace{.4pt}1007\discretionary{/}{%
}{/}s12650\discretionary{%
}{-}{-}021\discretionary{%
}{-}{-}00755\discretionary{%
}{-}{-}1}}


\bibitem{TA:Shusen:2012:GMMdvr}
S.~Liu, J.~A. Levine, P.-T. Bremer, and V.~Pascucci.
\newblock Gaussian mixture model based volume visualization.
\newblock In {\em IEEE Symposium on Large Data Analysis and Visualization (LDAV)}, pp. 73--77, Oct. 2012. doi: {{%
10\hspace{.1pt}\discretionary{.}{%
}{.}\hspace{.4pt}1109\discretionary{/}{%
}{/}LDAV\hspace{.1pt}\discretionary{.}{%
}{.}\hspace{.4pt}2012\hspace{.1pt}\discretionary{.}{%
}{.}\hspace{.4pt}6378978}}


\bibitem{TA:Lodha:1996:flowVisUncertainty}
S.~Lodha, A.~Pang, R.~Sheehan, and C.~Wittenbrink.
\newblock Uflow: visualizing uncertainty in fluid flow.
\newblock In {\em Proceedings of Seventh Annual IEEE Visualization '96}, pp. 249--254, Oct.-Nov. 1996. doi: {{%
10\hspace{.1pt}\discretionary{.}{%
}{.}\hspace{.4pt}1109\discretionary{/}{%
}{/}VISUAL\hspace{.1pt}\discretionary{.}{%
}{.}\hspace{.4pt}1996\hspace{.1pt}\discretionary{.}{%
}{.}\hspace{.4pt}568116}}


\bibitem{Lorensen:1987:MCA}
W.~E. Lorensen and H.~E. Cline.
\newblock Marching cubes: A high resolution {3D} surface construction algorithm.
\newblock {\em SIGGRAPH Computer Graphics}, 21(4):163--169, Aug. 1987. doi: {{%
10\hspace{.1pt}\discretionary{.}{%
}{.}\hspace{.4pt}1145\discretionary{/}{%
}{/}37402\hspace{.1pt}\discretionary{.}{%
}{.}\hspace{.4pt}37422}}


\bibitem{TA:Lundstrom:2007:probabilisticAnimationMedicalVolRendering}
C.~Lundstr\"{o}m, P.~Ljung, A.~Persson, and A.~Ynnerman.
\newblock Uncertainty visualization in medical volume rendering using probabilistic animation.
\newblock {\em IEEE Transactions on Visualization and Computer Graphics}, 13(6):1648--1655, Nov.-Dec. 2007. doi: {{%
10\hspace{.1pt}\discretionary{.}{%
}{.}\hspace{.4pt}1109\discretionary{/}{%
}{/}TVCG\hspace{.1pt}\discretionary{.}{%
}{.}\hspace{.4pt}2007\hspace{.1pt}\discretionary{.}{%
}{.}\hspace{.4pt}70518}}


\bibitem{TA:Otto:2011:3dVectorFieldTopologyUncertainty}
M.~Otto, T.~Germer, and H.~Theisel.
\newblock Uncertain topology of 3{D} vector fields.
\newblock In {\em 2011 IEEE Pacific Visualization Symposium}, pp. 67--74, Mar. 2011. doi: {{%
10\hspace{.1pt}\discretionary{.}{%
}{.}\hspace{.4pt}1109\discretionary{/}{%
}{/}PACIFICVIS\hspace{.1pt}\discretionary{.}{%
}{.}\hspace{.4pt}2011\hspace{.1pt}\discretionary{.}{%
}{.}\hspace{.4pt}5742374}}


\bibitem{visWithoutMean}
K.~Potter, S.~Gerber, and E.~W. Anderson.
\newblock Visualization of uncertainty without a mean.
\newblock {\em IEEE Computer Graphics and Applications}, 33(1):75--79, 2013. doi: {{%
10\hspace{.1pt}\discretionary{.}{%
}{.}\hspace{.4pt}1109\discretionary{/}{%
}{/}MCG\hspace{.1pt}\discretionary{.}{%
}{.}\hspace{.4pt}2013\hspace{.1pt}\discretionary{.}{%
}{.}\hspace{.4pt}14}}


\bibitem{TA:Potter:2012:UQtaxonomy}
K.~Potter, P.~Rosen, and C.~R. Johnson.
\newblock From quantification to visualization: A taxonomy of uncertainty visualization approaches.
\newblock In A.~M. Dienstfrey and R.~F. Boisvert, eds., {\em Uncertainty Quantification in Scientific Computing}, pp. 226--249. Springer Berlin Heidelberg, Berlin, Heidelberg, 2012. doi: {{%
10\hspace{.1pt}\discretionary{.}{%
}{.}\hspace{.4pt}1007\discretionary{/}{%
}{/}978\discretionary{%
}{-}{-}3\discretionary{%
}{-}{-}642\discretionary{%
}{-}{-}32677\discretionary{%
}{-}{-}6\_15}}


\bibitem{TA:Kai:2013:nonparametricIsoVis}
K.~Pöthkow and H.-C. Hege.
\newblock Nonparametric models for uncertainty visualization.
\newblock {\em Computer Graphics Forum}, 32(3pt2):131--140, July 2013. doi: {{%
10\hspace{.1pt}\discretionary{.}{%
}{.}\hspace{.4pt}1111\discretionary{/}{%
}{/}cgf\hspace{.1pt}\discretionary{.}{%
}{.}\hspace{.4pt}12100}}


\bibitem{featureConfidenceLevelSets}
S.~Sane, T.~M. Athawale, and C.~R. Johnson.
\newblock {Visualization of Uncertain Multivariate Data via Feature Confidence Level-Sets}.
\newblock In M.~Agus, C.~Garth, and A.~Kerren, eds., {\em EuroVis 2021 - Short Papers}. The Eurographics Association, 2021. doi: {{%
10\hspace{.1pt}\discretionary{.}{%
}{.}\hspace{.4pt}2312\discretionary{/}{%
}{/}evs\hspace{.1pt}\discretionary{.}{%
}{.}\hspace{.4pt}20211053}}


\bibitem{Shannon1948}
C.~E. Shannon.
\newblock A mathematical theory of communication.
\newblock {\em Bell System Technical Journal}, 27(3):379--423, 1948.

\bibitem{TA:Siddiqui:2021:tensorImagingVisPipelineUncertainty}
F.~Siddiqui, T.~Höllt, and A.~Vilanova.
\newblock A progressive approach for uncertainty visualization in diffusion tensor imaging.
\newblock {\em Computer Graphics Forum}, 40(3):411--422, June 2021. doi: {{%
10\hspace{.1pt}\discretionary{.}{%
}{.}\hspace{.4pt}1111\discretionary{/}{%
}{/}cgf\hspace{.1pt}\discretionary{.}{%
}{.}\hspace{.4pt}14317}}


\bibitem{vitart2017subseasonal}
F.~Vitart, C.~Ardilouze, A.~Bonet, A.~Brookshaw, M.~Chen, C.~Codorean, M.~D{\'e}qu{\'e}, L.~Ferranti, E.~Fucile, M.~Fuentes, et~al.
\newblock The subseasonal to seasonal (s2s) prediction project database.
\newblock {\em Bulletin of the American Meteorological Society}, 98(1):163--173, 2017.

\bibitem{TA:Whitaker:2013:contourBoxPlots}
R.~T. Whitaker, M.~Mirzargar, and R.~M. Kirby.
\newblock Contour boxplots: A method for characterizing uncertainty in feature sets from simulation ensembles.
\newblock {\em IEEE Transactions on Visualization and Computer Graphics}, 19(12):2713--2722, Dec. 2013. doi: {{%
10\hspace{.1pt}\discretionary{.}{%
}{.}\hspace{.4pt}1109\discretionary{/}{%
}{/}TVCG\hspace{.1pt}\discretionary{.}{%
}{.}\hspace{.4pt}2013\hspace{.1pt}\discretionary{.}{%
}{.}\hspace{.4pt}143}}


\bibitem{Zhan2014}
P.~Zhan, A.~C. S.~F. Yao, and I.~Hoteit.
\newblock Eddies in the {Red Sea}: A statistical and dynamical study.
\newblock {\em JGR Oceans}, (6):3909--3925, June 2014. doi: {{%
10\hspace{.1pt}\discretionary{.}{%
}{.}\hspace{.4pt}1002\discretionary{/}{%
}{/}2013JC009563}}


\bibitem{scatterPlotUncertainty}
B.~Zheng and F.~Sadlo.
\newblock Uncertainty in continuous scatterplots, continuous parallel coordinates, and fibers.
\newblock {\em IEEE Transactions on Visualization and Computer Graphics}, 27(2):1819--1828, 2021. doi: {{%
10\hspace{.1pt}\discretionary{.}{%
}{.}\hspace{.4pt}1109\discretionary{/}{%
}{/}TVCG\hspace{.1pt}\discretionary{.}{%
}{.}\hspace{.4pt}2020\hspace{.1pt}\discretionary{.}{%
}{.}\hspace{.4pt}3030466}}


\end{thebibliography}
\end{document}